\documentclass[%
 aip,
 amsmath,amssymb,
 reprint,%
]{revtex4-1}

\usepackage{amsmath}
\usepackage{amssymb}
\usepackage{wasysym}
\usepackage{graphicx}
\usepackage{hyperref}
\usepackage{color}
\usepackage{siunitx}
\usepackage{xcolor}
\usepackage{changes}
\usepackage{comment}
\usepackage{bm}
\usepackage{booktabs}
\usepackage{makecell}
\usepackage{xr}
\usepackage{booktabs}
\usepackage{diagbox}
\usepackage{multirow}
\usepackage{orcidlink}

\usepackage{ulem}
\AtBeginDocument{%
  \hypersetup{%
    colorlinks=true, linkcolor=blue,
    citecolor=blue, urlcolor=blue,
    breaklinks=true
  }%
}
\hypersetup{colorlinks=True,  linkcolor=blue,  citecolor=blue,   breaklinks}

\allowdisplaybreaks

\begin{document}

\title{A widely-tunable printed-circuit-board-based resonant electro-optic modulator}
\author{Jia-Xiao~Su}

\author{Xiang-Chuan~Yan}

\author{Hai-Xia~Li} 
\email[]{E-Mail: hxlee@hainanu.edu.cn}

\affiliation{School of Physics and Optoelectronic Engineering, Hainan University, 570228 Haikou, China}

\date{\today}

\begin{abstract}
We present a simple, printed-circuit-board (PCB)-based resonant electro-optic modulator (EOM) featuring a wide operational frequency range.  By utilizing a low-loss, high-frequency variable capacitor, the resonant frequency can be continuously tuned by typically  30\% within a single band, e.g.,  from 161.0 MHz to 233.5 MHz. Furthermore, by swapping a radio-frequency transformer, the EOM's resonance is reconfigurable across the 17.7--489.7 MHz range, with quality factors ranging from 46 to 109. Using this design, we demonstrate the EOM’s applicability by generating strong 228 MHz sidebands to verify its utility for $^{6}\text{Li}$ laser cooling and by locking a 780 nm laser to the Rb $\text{D}_2$ transition using retroreflected modulation transfer spectroscopy at 20 MHz resonance. The device's reconfigurable wide frequency range and high quality factors make this EOM versatile for laser cooling of multiple atomic species, including $^{6}\text{Li}$, $^{39}\text{K}$, and $^{41}\text{K}$, as well as for laser frequency locking. This design offers a flexible, robust, and practical solution for undergraduate teaching laboratories and quantum optics research groups.

\end{abstract}

\maketitle
\vspace{0.5em}
\noindent\rule{\linewidth}{0.4pt} 
\begin{flushleft}
\small
\textbf{Accepted Manuscript}\\
\copyright\ 2026 American Institute of Physics. This is the accepted manuscript of the following article: Jia-Xiao Su, Xiang-Chuan Yan, and Hai-Xia Li, ``A widely tunable printed-circuit-board-based resonant electro-optic modulator,'' Review of Scientific Instruments, 97(11), 113201 (2026). The final published version is available at \href{https://doi.org/10.1063/5.0354726}{https://doi.org/10.1063/5.0354726}.
\end{flushleft}
\vspace{-0.5em}
\noindent\rule{\linewidth}{0.4pt} 
\vspace{1em}

\section{Introduction}\label{Introduction}

Laser cooling and trapping are widely used in atomic and molecular physics, precision measurements, and quantum research. \cite{Wineland1979,chu1985,Bloch2008,Bakr2009,Ludlow2015,Langen2024} In these experiments, lasers often need to be locked to atomic transitions, or multiple frequency components must be generated to address different hyperfine levels. \cite{Hemmerling2011,Chepurov2014}
Electro-optic modulators (EOMs) are key devices for laser phase modulation and are widely employed in laser cooling of atoms and molecules \cite{Kelly1987} and in laser frequency stabilization \cite{Drever1983,Black2001} . They are also used for phase control of squeezed light in quantum optics \cite{Suerra2026}.

The required modulation frequencies vary significantly across applications. For laser frequency locking using modulation transfer spectroscopy (MTS) \cite{McCarron2008} or Pound–Drever–Hall (PDH) \cite{Black2001} techniques, a few tens of MHz is sufficient to produce a clean error signal. In contrast, direct laser cooling of atoms like \(^{6}\text{Li}\) and \(^{39}\text{K}\) requires sidebands that match their ground-state hyperfine splittings, which are hundreds of MHz (e.g., 228 MHz for \(^{6}\text{Li}\) \cite{Schlecht1966,Walls2003}, 461 MHz for \(^{39}\text{K}\) \cite{Kusch1940,Peper2019}). Resonant EOMs are preferred for these tasks because they offer high modulation depth with moderate radio-frequency (RF) drive power, however, most existing resonant designs are fixed-frequency, making them inflexible for laboratories that handle both locking and cooling of multiple species (e.g., Li–K mixtures). Although commercially tunable EOMs are available \cite{thorlabs_eo_pm_nr_c1,qubig_eom,lbtek_eopm_20m_a}, their internal designs are proprietary, making customization or repair difficult. On the other hand, home-built resonant EOMs reported in the literature are often designed for fixed frequencies in the tens of MHz range \cite{tian2023,Izumi2012,Dooley2011}, and thus cannot generate the high-frequency sidebands needed for lithium or potassium cooling.

To overcome these limitations, we present a simple printed-circuit-board-based resonant EOM with a modular LC circuit design.\cite{patent1993} Our device achieves a wide reconfigurable frequency range by using interchangeable RF transformers for band selection, combined with an on-board variable capacitor for continuous fine-tuning within each band. We test its phase modulation performance and characterize its quality factors across multiple frequency bands. To demonstrate its practical use, we apply the EOM to generate 228 MHz sidebands at 780 nm to verify its capability for $^{6}\text{Li}$ laser cooling, and to stabilize 
a rubidium laser via a retroreflected modulation transfer spectroscopy configuration. The compact, open layout and the use of standard components make it easy to assemble and replicate.

The rest of this paper is organized as follows. Section II describes the circuit principle and mechanical design. Section III presents the RF characterization and optical sideband generation results, and shows the laser frequency locking performance. Section IV discusses the experimental results. Section V concludes the paper.

\begin{figure}
    \centering
    \includegraphics[width=1\linewidth]{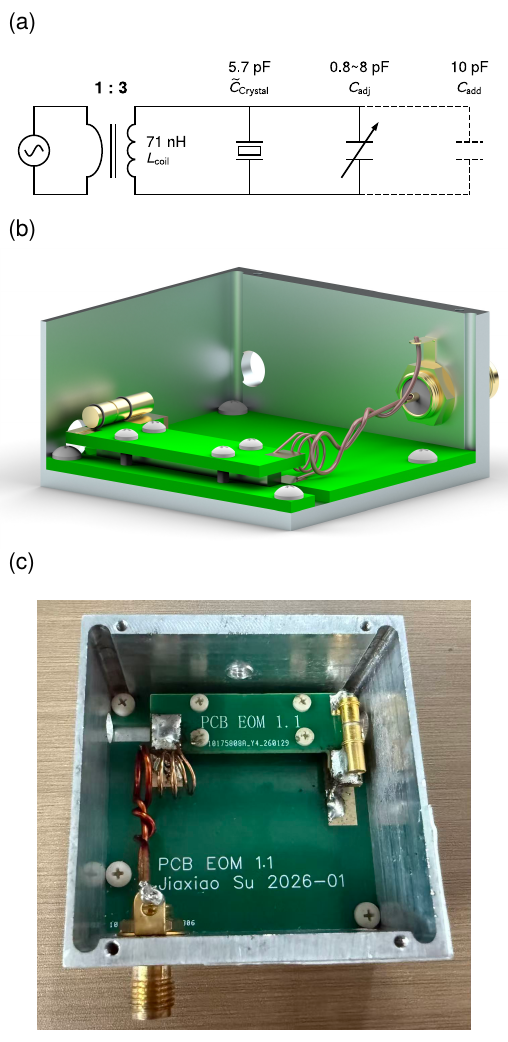}
    \caption{ Schematic and mechanical design of the home-built PCB-based resonant EOM (228 MHz).  (a) Equivalent circuit diagram of the LC resonant circuit,  which comprises an on-board RF transformer, a modular inductor coil ($L_{\text{coil}}$), a low-loss variable capacitor ($C_{\text{adj}}$), the effective crystal capacitance ($\tilde{C}_{\text{crystal}}$) and optional additional capacitance ($C_{\text{add}}$).  (b) 3D mechanical model of the assembled EOM.  The lithium niobate ($\text{LiNbO}_3$) crystal is clamped between two custom PCBs and enclosed in a compact ($45 \times 46 \times 25\text{ mm}$) aluminum shielding box.  (c) Photograph of the assembled 228 MHz PCB-based resonant EOM corresponding to the model in (b).}
    \label{FIG.  1. Schematic diagram}
\end{figure}
\section{Principle and  Design}\label{Principle and Design}

As shown in Fig.~1(a), this EOM is built around a resonant LC circuit \cite{patent1993} on two simple PCBs. The circuit consists of a primary coil and a secondary resonant circuit, which together form an RF transformer. This secondary part consists of an inductor coil, an electro-optic crystal, and an adjustable capacitor. Here, the lithium niobate (LiNbO$_3$) crystal acts as the main capacitive load ($\tilde{C}_{\mathrm{crystal}}$ in Fig.~1(a)). To tune the frequency precisely, we connect a low-loss, high-frequency variable capacitor ($C_{\mathrm{adj}}$, tuning range 0.8--8~pF) in parallel with the crystal. A transformer is integrated into the design for efficient impedance matching. This transformer couples the $50~\Omega$ RF source to the resonant tank, 
matching the input impedance to the high equivalent parallel resistance of the circuit at the operating frequency to minimize RF reflections. At resonance, 
the high quality factor $Q$ of the parallel LC loop ensures that a substantial RF voltage is established across the $\text{LiNbO}_3$ crystal for a given input power. This combined impedance matching and resonant enhancement enables a high modulation depth even with moderate RF driving power. The resonant frequency of this resonant circuit can be expressed approximately as:
\begin{equation}
f_0 = \frac{1}{2\pi\sqrt{L (\tilde{C}_{\mathrm{crystal}}+C_{\mathrm{add}}+C_{\mathrm{adj}})}} .
\label{eq:resonance}
\end{equation}
Here, \(f_0\) is determined by the total effective inductance \(L\), the effective capacitance of the LiNbO$_3$ crystal capacitance \(\tilde{C}_{\text{crystal}}\), the adjustable capacitance \(C_{\text{adj}}\), and an optional additional capacitance \(C_{\text{add}}\). A quantitative estimate of effective crystal capacitance $\tilde{C}_{\text{crystal}}$ and effective inductance $L$ is given in Appendix~\ref{app:parasitics} and summarized in Table~\ref{tab:tab4}. 
To meet the requirements of different experiments, the resonant frequency can be reconfigured by replacing the RF transformer module or be finely tuned within a band using the variable capacitor $C_{\text{adj}}$. Tuning $C_{\text{adj}}$ provides continuous fine tuning of typically 30\% without changing the transformer, whereas replacing the module with different turns ratios and secondary coil geometries shifts the resonance to a different frequency range. The primary coil is also optimized per band to maintain efficient impedance matching, while the secondary coil primarily determines the resonant frequency; its parameters for each band are listed in Table~\ref{tab:tab1}. For low-frequency operation (e.g., 20 MHz), an additional 10 pF high Q ceramic capacitor is connected in parallel with $\tilde{C}_{\text{crystal}}$ and $C_{\text{adj}}$ to shift the resonance to the target low-frequency range.

The mechanical design of the device is shown in Fig.~1(b). The entire EOM is housed in an aluminum alloy shielding box with dimensions of $45 \times 46 \times 25$~mm. All components are mounted on the PCBs to shorten the interconnection lengths, thereby reducing parasitic inductance in the high-frequency circuit. The PCBs are mounted to the bottom of the aluminum box using PEEK screws, ensuring the resonant frequency remains stable against vibrations. We used a $3 \times 3 \times 25\text{~mm}$ $x$-cut $\text{LiNbO}_3$ crystal clamped between the two custom PCBs. The $\text{LiNbO}_3$ crystal is coated with anti-reflection coating on the $x$-faces for the 500-900 nm spectral range, and with gold on the $z$-faces. We aligned its optical axis with the holes on the sides of the box so that the laser beam can pass through the crystal and be modulated by the electric field. A 5~mm tuning hole on the aluminum box allows users to adjust the capacitor with a screwdriver without opening the enclosure. This design results in a compact EOM structure that is mechanically stable and relatively easy to assemble. For ease of replication, the main components and their suppliers are summarized in Table~\ref{tab:tab2}.

\begin{table}[h]
\caption{\label{tab:tab1}
A table summarizing the parameters of the EOMs characterized in Fig.~4, including the frequency tuning range, winding number $N_{\mathrm{coil}}$, diameter $d_{\mathrm{coil}}$, length $l_{\mathrm{coil}}$, and wire thickness $t_{\mathrm{coil}}$ of the secondary coil. The primary coil has one turn for most bands (two for 17.7--20.7 MHz band) and a diameter equal to or slightly larger than that of the secondary coil.
}
\begin{ruledtabular}
\begin{tabular}{lcccc}
Frequency (MHz) & $N_{\mathrm{coil}}$ & $d_{\mathrm{coil}}$ (mm) & $l_{\mathrm{coil}}$ (mm) & $t_{\mathrm{coil}}$ (mm) \\
$17.7\text{--}20.7$\footnote{An additional 10 pF ceramic capacitor is connected in parallel with $\tilde{C}_{\mathrm{crystal}}$ and $C_{\mathrm{adj}}$ to lower its resonant frequency.} & 22 & 14 & 24 & 0.8 \\
$50.0\text{--}67.4$   & 10 & 14 & 24  & 1   \\
$95.7\text{--}135.6$  & 3  & 14 & 4   & 1   \\
$161.0\text{--}233.5$ & 3  & 6  & 5   & 0.8 \\
$247.3\text{--}407.5$ & 1  & 6  & 0.8 & 0.8 \\
$286.8\text{--}489.7$ & 1  & 3  & 0.8   & 0.8\
\end{tabular}
\end{ruledtabular}

\end{table}

\begin{table*}[t] 
\caption{\label{tab:tab2}
This table summarizes the quantity, model, and
supplier of the main components required to assemble a
home-built EOM.
}
\begin{ruledtabular}
\begin{tabular}{llcc}
Main components & Quantity & Part number & Supplier \\
Variable capacitor & 1 & 843-27291SL & Mouser Electronics \\
Upper PCB (FR-4) & 1 & custom design  & JLC Technology Group \\
Bottom PCB (FR-4) & 1 & custom design & JLC Technology Group \\
$x$-cut $\text{LiNbO}_3$ crystal & 1 & LN-3325 & MT Optics \\
SMA connector & 1 & SMA-KY & Shenzhen Lianfaxin Electronic Technology Enterprise\\
Aluminum shielding box & 1 & custom design & Shenzhen Jianye Plastic and Hardware Products Factory\\
Inductor coil & 2 & custom design & Front Hardware Flagship Store \\
\end{tabular}
\end{ruledtabular}
\end{table*}

\section{Characterization}\label{Characterization}

To evaluate the phase modulation capability of the PCB-based resonant EOM, we sent a 780 nm laser beam through the EOM operating at a resonant frequency of 228~MHz. The phase-modulated optical spectrum was characterized using a scanning Fabry--Pérot interferometer (FPI). The transmitted signal from the FPI was detected by a photodetector and recorded on an oscilloscope.

\begin{figure}
    \centering
    \includegraphics[width=1\linewidth]{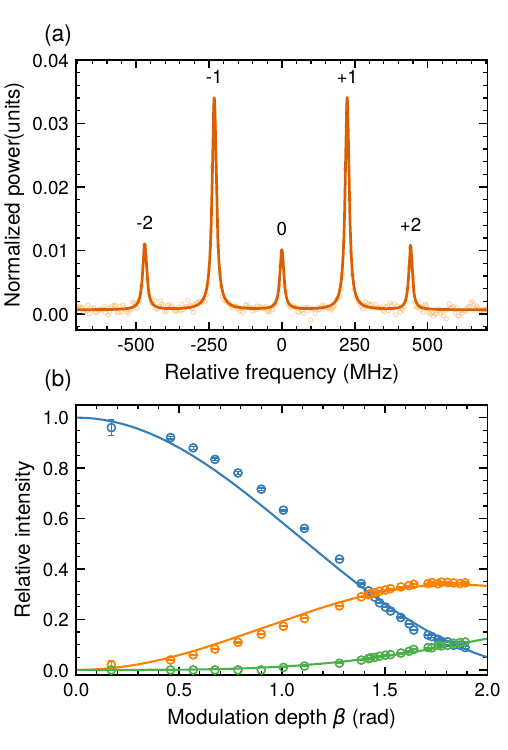}
    \caption{Optical sideband spectrum and Bessel-function fitting. (a) FPI transmission spectrum of the carrier and first- and second-order sidebands generated by the 228~MHz EOM at 30.89~dBm ($\beta = 1.85 \pm 0.02$). (b) Normalized powers of the carrier (blue), first-order sideband (orange), and second-order sideband (green) versus $\beta$. Solid curves are Bessel-function fits, yielding $V_{\pi} = 19.0 \pm 0.1~\mathrm{V}$. The carrier and first-order sideband powers are equal at $\beta = 1.4347$. Error bars in (b) represent propagated $1\sigma$ uncertainties from the covariance matrices of the multi-Lorentzian fits.}
    \label{FIG.  2.  Bessel function fitting curve}
\end{figure}

\begin{figure}
    \centering
    \includegraphics[width=\linewidth]{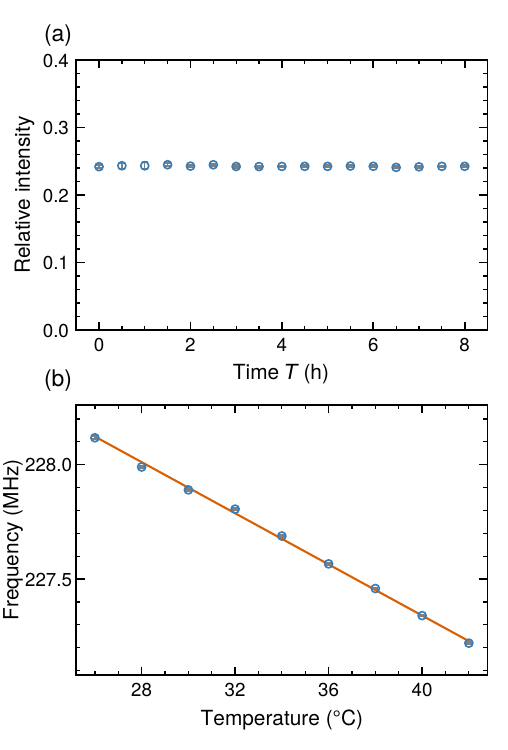}
    \caption{Long-term sideband stability and temperature dependence of the
nominally 228~MHz EOM.
(a) Normalized first-order sideband intensity measured over 8~h
at an RF input power of 27.3~dBm.
Following a 30-min warm-up, spectra were recorded at 30-min
intervals with the ambient temperature maintained at
$22.0 \pm 0.1\,^{\circ}\mathrm{C}$.
The mean relative intensity is approximately 24\%, with a
relative root-mean-square (RMS) deviation of 0.39\%.
Error bars represent propagated $1\sigma$ uncertainties
from the multi-Lorentzian fits.
(b) Resonant frequency as a function of the measured
ambient temperature $T$ over the range of
26--42~$^{\circ}\mathrm{C}$.
Blue open circles show the measured resonant frequencies
with $1\sigma$ error bars, and the orange solid line is
a weighted linear least-squares fit, yielding
$\mathrm{d}f_0/\mathrm{d}T =
(-55.90 \pm 0.29)\,\mathrm{kHz}/^{\circ}\mathrm{C}$.
The quoted uncertainty is the $1\sigma$ fit uncertainty.}
    \label{fig:temperature_dependence}
\end{figure}

Fig.~2(a) displays the measured FPI transmission spectrum, showing the central carrier along with symmetrically distributed first-order and second-order sidebands resulting from the phase modulation. A laser beam with a carrier frequency $\omega_L$ undergoes phase modulation upon passing through an electro-optic crystal driven by an electric field at modulation frequency $\Omega$. The modulation depth $\beta$ characterizes the strength of the phase modulation and represents the maximum phase shift experienced by the laser beam as it passes through the crystal. $\beta$ is related to the peak-to-peak voltage of the input signal $V_{\mathrm{pp}}$ by $\beta=\pi V_{\mathrm{pp}}/(2V_\pi)$, where $V_\pi$ is the half-wave voltage of the EOM, defined as the voltage across the crystal required to induce a phase shift of $\pi$. The complex amplitude of the output optical field $E(t)$ can be described as: \cite{Kawanishi2022}
\begin{equation}
E(t)=E_0\sum_{n=-\infty}^{\infty}J_n(\beta)e^{i(\omega_L+n\Omega)t}.
\end{equation}

Here,  Eq.~(2) indicates that the original carrier signal is modulated into a spectrum consisting of a central carrier ($n=0$) and infinitely many symmetrically distributed sidebands (n=$\pm$1, $\pm$2, \ldots  ).  $J_n(\beta)$ is the n-th order Bessel function of the first kind,  $E_0$ is the incident light amplitude. Since photo detectors measure optical intensity, which is proportional to the squared modulus of the complex field amplitude, the optical power in the n-th order component is proportional to $\left|J_n(\beta)\right|^2$. In the data analysis, the peak heights of the carrier and the first- and second-order sidebands were extracted from the FPI spectra at different RF driving powers. For each spectrum, the ratios of the first and second order sideband peak heights to the carrier peak height were determined. The carrier and the first and second order sidebands were then normalized to obtain the relative intensities shown in Fig.~2(b). For each RF driving power, the corresponding peak-to-peak voltage $V_{\mathrm{pp}}$ was calculated assuming an input impedance of $50~\Omega$. For a given value of $V_\pi$, the corresponding modulation depth $\beta$ was calculated using $\beta=\pi V_{\mathrm{pp}}/(2V_\pi)$. The normalized relative intensities of the carrier, first order sidebands, and second-order sidebands were then fitted simultaneously with the corresponding Bessel function terms $J_0^2(\beta)$, $J_1^2(\beta)$, and $J_2^2(\beta)$ respectively. In this global fit, $V_\pi$ was treated as a shared fitting parameter for all three components. The fit yields a half-wave voltage of $V_\pi = 19.0 \pm 0.1~\mathrm{V}$ at $780~\text{nm}$.

It is worth noting that while this characterization was performed using a \(780~\mathrm{nm}\) laser, the actual laser cooling of \(^{6}\mathrm{Li}\) requires a wavelength of \(671~\mathrm{nm}\). Since both the refractive index and the relevant electro-optic coefficient of LiNbO\(_3\) are wavelength-dependent, the half-wave voltage also depends on wavelength. The wavelength dependence of the half-wave voltage can be described by \(V_\pi \propto \lambda d/(n^3 r L)\) \cite{Saleh2019}, where \(\lambda\) is the optical wavelength, \(n\) and \(r\) are the relevant refractive index and electro-optic coefficient, \(d\) and \(L\) are the electrode spacing and interaction length. For a fixed transverse bulk (non-waveguide) device geometry the geometric factor \(d/L\) is constant. Hence, \(V_\pi\) scales approximately as \(\lambda/(n^3 r)\). Thus, a shorter wavelength directly reduces \(V_\pi\) through the linear \(\lambda\) dependence. In addition, since LiNbO\(_3\) exhibits normal dispersion, \(n(671~\mathrm{nm}) > n(780~\mathrm{nm})\), which further lowers \(V_\pi\) at \(671~\mathrm{nm}\). The wavelength dependence of \(r\) is relatively weak in the visible range and does not change the overall trend. Therefore, theoretically, \(V_\pi\) is smaller at \(671~\mathrm{nm}\) than at \(780~\mathrm{nm}\), which, under otherwise identical conditions, can favor stronger modulation sidebands for \(^{6}\mathrm{Li}\) laser cooling at the same RF driving power.

According to the fitted Bessel-function curves, the carrier and each of the first-order sidebands have equal relative intensities at $\beta = 1.4347$, corresponding to a driving power of $28.8~\mathrm{dBm}$.
This result shows that strong first order sidebands can be generated with moderate driving power. Owing to the voltage enhancement provided by the LC resonant circuit, the EOM can achieve deep modulation with a standard laboratory signal generator, either directly or with a simple low-power amplifier. This removes the need for a high-power RF amplifier, reducing both the size and cost of the system. 

To evaluate the operational stability of the EOM, we monitored the normalized first-order sideband intensity over 8~h at a modulation frequency of 228~MHz and an RF input power of
27.3~dBm. After a 30-min warm-up, FPI spectra were recorded at 30-min intervals while the ambient temperature was maintained at
$22.0 \pm 0.1\,^{\circ}\mathrm{C}$. As shown in Fig.~3(a), the normalized first-order sideband intensity remained at a mean value of approximately 24\%, with a relative RMS deviation of 0.39\%. These results indicate that the modulation performance remained stable during the measurement period. To quantify the sensitivity of the resonance to ambient temperature, we measured the resonant frequency of the nominally 228~MHz EOM as the ambient temperature was varied from 26 to 42~$^{\circ}\mathrm{C}$. As shown in Fig.~3(b), the resonant frequency decreases approximately linearly with ambient temperature. A weighted linear least-squares fit yields
$\mathrm{d}f_0/\mathrm{d}T =
(-55.90 \pm 0.29)\,\mathrm{kHz}/^{\circ}\mathrm{C}$,
where $T$ denotes the measured ambient temperature. The EOM resonant frequency is extremely insensitive to typical laboratory ambient-temperature fluctuations.
For example, an ambient-temperature change of
$2\,^{\circ}\mathrm{C}$ produces a resonance shift of approximately 112~kHz at 228~MHz, corresponding to a fractional change of only
$|\Delta f_0|/f_0 = 4.9\times10^{-4}$
(approximately $1/2000$). This frequency shift is much smaller than the measured resonance linewidth of approximately 2~MHz.
Consequently, such ambient-temperature fluctuations are expected to have a negligible effect on the sideband intensity when the EOM is operated near resonance.
The 30-min warm-up preceding the measurement in Fig.~3(a) primarily allows the crystal to approach thermal equilibrium under RF heating.
Following this warm-up, the measured sideband intensity exhibits excellent stability over 8~hours.

\begin{figure}
    \centering
    \includegraphics[width=1\linewidth]{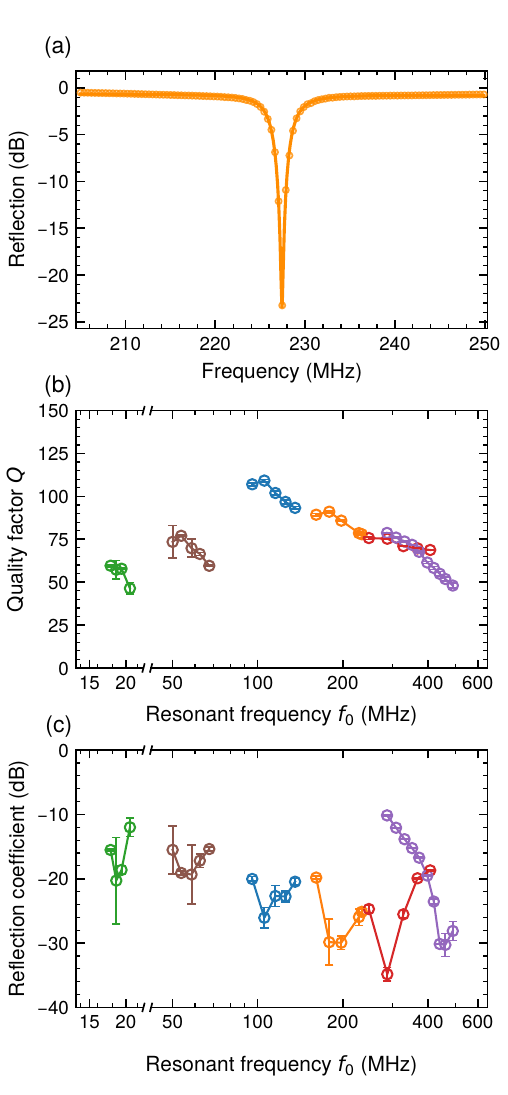}
    \caption{RF characterization and frequency tuning of the PCB-based resonant EOM.
(a) Measured reflection spectrum around the 228~MHz resonance, showing a resonance dip of approximately $-23$~dB.
(b) Quality factor $Q$ as a function of the resonant frequency $f_0$ for six interchangeable RF transformer modules, covering a frequency range from 17 to 490~MHz. The colors denote the coil configurations listed in Table~I. The error bars represent the combined $1\sigma$ uncertainty in the mean $Q$, including the standard error from repeated measurements and the fitting uncertainty propagated from $f_0$ and $\gamma$.
(c) On-resonance reflection coefficient $S_{11}$ across the corresponding frequency bands. The error bars represent the standard error of the mean on-resonance reflection coefficient obtained from repeated measurements.}
    \label{FIG. 3. Characterization of the RF resonance.  }
\end{figure}

\begin{figure}[!t]
    \centering
    \includegraphics[width=1\linewidth]{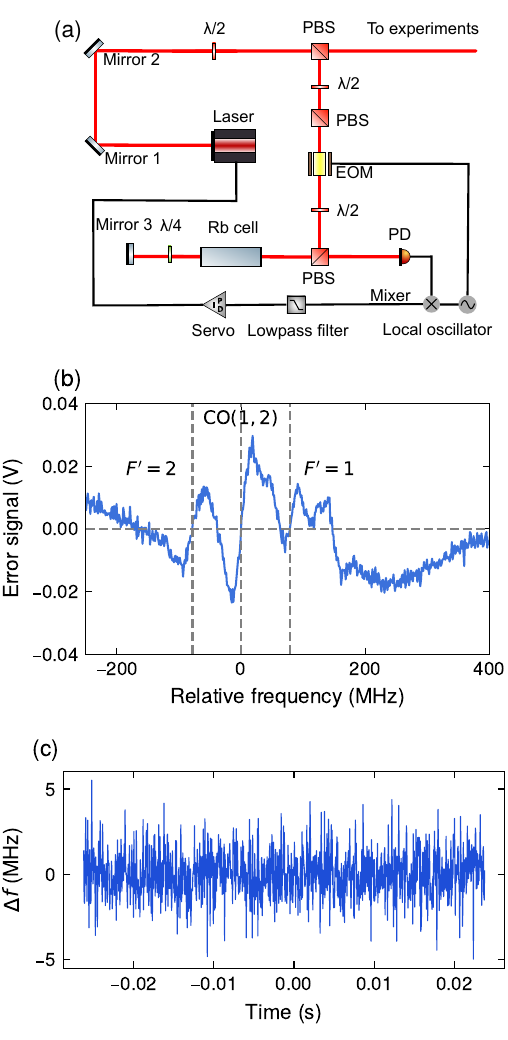}
    \caption{Laser frequency locking performance. (a) Optical setup for laser frequency locking via a retroreflected MTS configuration. PBS denotes a polarizing beam splitter, while $\lambda/2$ and $\lambda/4$ denote half-wave and quarter-wave plates, respectively. (b) Error signal from the $^{87}$Rb $\text{D}_2$ $F=1 \rightarrow F'$ transitions. From left to right, dashed lines indicate $F=1$ to $F'=2$, $F'=\text{CO}(1,2)$, and $F'=1$  resonances, respectively. The laser is locked to the zero-crossing of the $F'=\text{CO}(1,2)$ transition. (c) Frequency jitter when the laser is locked to the $F'=\text{CO}(1,2)$ crossover.}
    \label{fig:placeholder}
\end{figure}

To characterize the energy coupling efficiency of our PCB EOM,  we used a vector network analyzer (ZVB8, Rohde \& Schwarz) to measure the reflection coefficient ($S_{11}$) of the RF signal loaded onto the crystal. Fig. 4(a) illustrates the measured $S_{11}$ spectrum around $f_0 \approx 228$ MHz, which corresponds to the ground-state hyperfine splitting of $^{6}\text{Li}$ atoms. We observe a pronounced resonant reflection dip, with a reflection coefficient $-23$ dB. The low reflection coefficient indicates that the RF signal can couple efficiently into the EOM load. To determine the resonance parameters, the measured $S_{11}$ spectrum was fitted with a complex Fano-Lorentzian reflection model. The resonant frequency $f_0$ and the resonance linewidth parameter $\gamma$ were extracted from the fit, and the quality factor was calculated as $Q=f_0/\gamma$.

Fig.~4(b) summarizes the resonant frequencies and the corresponding quality factors ($Q$) for the six different inductor coils. The parameters of each configuration are listed Table~\ref{tab:tab1}. The PCB EOM's resonant frequency can be reconfigured from 17.7~MHz to 489.7~MHz. Within each band, the variable capacitor provides $16\%$--$50\%$ continuous tuning. The Q factor ranges from 46 to 109 across the entire operating range, with a peak value of 109 in the 100~MHz band. The resonant LC circuit exploits the high quality factor $Q$ to significantly boost the voltage across the $\mathrm{LiNbO}_3$ crystal, thereby reducing the RF power required for a given modulation depth. For the $17.7 - 20.7\ \mathrm{MHz}$ band, the additional 10 pF capacitor yields a Q of approximately 50. At 20 MHz, this is higher than the standard commercial resonant EOM, which has a typical $Q \approx 15$ \cite{thorlabs_eo_pm}. In the high frequency range ($>400$~MHz), the Q factor remains above 50. These results confirm efficient impedance matching and high Q factors across the tuning range. Fig.~4(c) shows the on-resonance reflection coefficient $S_{11}$ for each frequency band. For all measured configurations, the on-resonance $S_{11}$ remains below $-10$~dB, corresponding to less than 10\% of the incident RF power reflected. Together, these results demonstrate effective impedance matching and consistently high $Q$ factors across the entire measured tuning range.

To demonstrate the use of the EOM for laser frequency stabilization, we implemented a retroreflected modulation transfer spectroscopy configuration (hereafter referred to as retroreflected MTS; the error signal contains contributions from both modulation transfer and frequency modulation) based on rubidium atomic transitions. Fig.~5(a) shows the schematic of the optical setup. A portion of a 780~nm laser (FECL-SF-780-S, Precilasers) is sent to the frequency-locking branch. In this path, the beam is phase-modulated at 20~MHz by the PCB EOM (configured for low-frequency operation, Table~\ref{tab:tab1}). After passing through a rubidium (Rb) vapor cell, the beam is retro-reflected to propagate back along the same path, forming a pump–probe configuration. The transmitted probe light is detected by a photo-detector. Since both beams carry modulation sidebands, the error signal arises from a mix of modulation transfer and frequency modulation \cite{McCarron2008}. Although this scheme leaves a small residual Doppler background (compared to MTS), it greatly simplifies the optics and reduces components. Demodulation of the detected signal yields the error signal shown in Fig.~5(b), which corresponds to the $\text{D}_2$ $F=1 \rightarrow F'$ transitions of the $^{87}$Rb atoms. Owing to EOM resonant enhancement, we obtain an error signal with a steep zero-crossing slope using a low RF driving power of about 8.3 dBm. This enables the feedback system to detect small frequency fluctuations and to feed a correction signal back to the injection current of the laser. The laser frequency is locked to the atomic transition using a commercial frequency-locking module (Precilock-PDH-HV, Precilasers). Fig.~5(c) shows the residual frequency fluctuations of the locked laser over time, obtained by converting the error-signal voltage fluctuations using the locally calibrated zero-crossing slope $\mathrm{d}V_{\mathrm{err}}/\mathrm{d}f$, yielding an RMS frequency deviation of $1.43~\mathrm{MHz}$, which indicates sufficient frequency stability for laser cooling of rubidium atoms.

\section{Discussion}

\begin{table*}[t]
\caption{\label{tab:tab3}
Comparison of typical commercial resonant electro-optic modulators and the
PCB-based resonant EOM developed in this work.}
\centering
\footnotesize
\setlength{\tabcolsep}{2pt}
\renewcommand{\arraystretch}{1.12}
\begin{ruledtabular}
\begin{tabular}{lcccccc}
Manufacturer &
Model &
Frequency range (MHz) &
$Q$ &
$V_{\pi}$  &
Dimensions (mm) &
Internal design\\[3pt]

Thorlabs &
EO-PM-R-20-C1 ~\cite{ThorlabsEOPMR20C1} &
20 fixed &
15 &
10 V @ 633 nm &
$66 \times 31.8 \times 31.8$ &
Proprietary \\

This work &
PCB EOM &
17.7--20.7 &
46 @ 20 MHz &
---\footnote{Not measured as the 20~MHz sideband can not be
resolved by our Fabry--Perot cavity. However, its \(Q\) is
notably larger than that of the commercial EOM.
Note that the frequency locking with the 20 MHz configuration was achieved with only
8.3~dBm RF power.}  &
$45 \times 46 \times 25$ &
Open source\\[3pt]

QUBIG &
PM-Li6\_0.2M3~\cite{QUBIGLi6EOM} &
196--265 &
255 @ 229 MHz &
11.5 V @ 671 nm &
$40 \times 40 \times 40$ &
Proprietary \\

This work &
PCB EOM &
161.0--233.5 &
78 @ 228 MHz &
$19.0 \pm 0.1$ V @ 780 nm &
$45 \times 46 \times 25$   &
Open source\\

\end{tabular}
\end{ruledtabular}
\end{table*}

Compared with conventional resonant EOMs operating at fixed frequencies, the present design offers the flexibility to switch between different resonant frequency bands by exchanging the RF transformer modules, while a variable capacitor enables continuous fine tuning within each band. This modular design allows the same PCB platform to accommodate different modulation-frequency requirements without redesigning the entire device for each operating frequency.

Notably, the device maintains favorable resonant characteristics as the coil and operating frequency are changed. Across the measured configurations, the quality factor ranges between approximately 46 and 109, indicating that the frequency-tuning approach maintains effective resonant voltage enhancement. At 20~MHz, our EOM achieves \(Q = 46\), higher than the \(Q \approx 15\) of the Thorlabs EO-PM-R-20-C1. At 228~MHz, our EOM achieves \(Q = 78\), while the QUBIG PM-Li6\_0.2M3 reaches \(Q = 255\) at 229~MHz.  Although our \(Q\) is lower than that of this higher-\(Q\) commercial device, it remains sufficient for efficient resonant enhancement and deep modulation with moderate RF drive power. The primary advantage of our design is therefore its broad tunability and open-source architecture, rather than a universally higher \(Q\) than all commercial devices. A detailed comparison with commercial resonant EOMs is summarized in Table~\ref{tab:tab3}.

Two trends in the $Q$ data merit further discussion. First, within each individual band, $Q$ tends to decrease as the resonant frequency is tuned upward via the variable capacitor. This behavior is likely due to the increased losses in both the tuning capacitor and the coil at higher frequencies. Second, across different bands, $Q$ exhibits a clear rising trend from low to moderate frequencies, i.e., from approximately 50 at 17.7--20.7 MHz to approximately 100 at 95.7--135.6 MHz, followed by a gradual decline at higher frequencies. We attribute this trend to the coil parameters listed in Table~\ref{tab:tab1}. The peak at 95.7--135.6 MHz (3 turns, 14 mm diameter) may reflect an optimal balance: the winding number is low enough to minimize resistive losses, yet the frequency is not so high that high-frequency losses dominate. At lower frequencies (17.7--20.7 MHz, 22 turns), the increased series resistance and inter-turn capacitance likely degrade $Q$ to $\sim$50. At higher frequencies ($>$400 MHz, single-turn coils), these losses, together with the parasitic capacitance of the FR-4 substrate, probably become more significant, pulling $Q$ back to approximately 60.

The device therefore combines band reconfiguration with resonant modulation. This tunability makes the EOM suitable both for multi-species laser cooling and for laser frequency locking. With the present PCB design, the maximum resonant frequency is limited to about 500 MHz by roughly 25 nH parasitic inductance, as discussed in Appendix~\ref{app:parasitics}. In future work, further optimization of the PCB layout and secondary coil design could potentially extend the resonant frequency into the GHz range. The quality factor and heat dissipation might also be improved by using ultralow-loss PCB materials with high thermal conductivity, such as aluminum nitride or silicon nitride substrates.

Although our current prototype uses a standard rectangular LiNbO$_3$ crystal, it can be replaced in the future with crystals of customized geometries, such as wedged or Brewster-angled cuts, or even combined with active temperature control to further suppress residual amplitude modulation (RAM) \cite{Li2021,Li2016,Tai2016}. The open layout of our PCB platform allows crystals to be swapped without modifying the core electronics, making the architecture readily adaptable to more demanding applications.

\section{Conclusion}\label{Conclusion}

In conclusion, we have developed a widely reconfigurable resonant EOM using a simple PCB design. In contrast to conventional fixed‑frequency versions, our EOM covers a broad range from tens to hundreds of MHz by swapping RF transformer modules, with an onboard variable capacitor for continuous fine‑tuning within each band. The low-loss circuit and efficient impedance matching result in a high-Q LC resonator and low half‑wave voltage, enabling deep phase modulation at hundreds of MHz with moderate driving power. Beyond laser locking, its multi-band flexibility supports laser cooling of various species. The robust sideband generation demonstrated here highlights its practical utility for multi-species ultracold atom experiments, particularly for $^{6}\mathrm{Li}$ \cite{Schlecht1966,Walls2003}, $^{39}\mathrm{K}$ \cite{Peper2019}, and $^{41}\mathrm{K}$ \cite{Kusch1940}. Moreover, the open and modular design makes the EOM easy to replicate and adapt, providing a flexible platform for both teaching and advanced research.

\begin{acknowledgments}
We gratefully acknowledge  Z. Ye and Y. Cui for stimulating discussions. This work is supported by the National Natural Science Foundation of China under Grants No. 12504312. We gratefully acknowledge the starting fund from Hainan University. A utility model patent (Application No. 2026214235852) has been filed based on the content presented in this manuscript.

\end{acknowledgments}
\section*{Author Declarations}
The authors declare that there are no conflicts of interest related to this work.
\section*{Data Availability Statement}
The design files are openly available in the Github repository~\cite{pcb_eom_git}. The data that support the findings of this study are available from the corresponding author upon reasonable request.

\bibliography{main}

\clearpage

\appendix
\section{Estimation of effective inductance and effective crystal capacitance}
\label{app:parasitics}

In high-frequency circuits, it's difficult to model the parasitic capacitance and inductance of the PCB layout, interconnects, and nearby components in a precise equivalent-circuit. In order to account for these effects within a simplify model, we absorb the effect of the parasitic capacitance into the effective crystal capacitance \(\tilde{C}_{\mathrm{crystal}}\) and the parasitic inductance into the total effective inductance $L$, and approximate the resonant tank as the secondary coil in parallel with \(\tilde{C}_{\mathrm{crystal}}\), the additional capacitance $C_{\mathrm{add}}$, and the adjustable capacitance \(C_{\mathrm{adj}}\). We experimentally determine $L$ and \(\tilde{C}_{\mathrm{crystal}}\) as follows: for each configuration we set the adjustable capacitance \(C_{\mathrm{adj}}\) to its maximum (8 pF) and minimum (0.8 pF) and record the corresponding resonant frequencies. Using the relation between the resonant frequency and the total capacitance, Eq.~\eqref{eq:resonance}, with $C_{\mathrm{add}}=10$ pF in the 17.7--20.7 MHz band and 0 pF in other bands, \(\tilde{C}_{\mathrm{crystal}}\) is determined to be \(9.3~\mathrm{pF}\) in the 17.7--20.7 MHz band and decreases to \(3.0~\mathrm{pF}\) in the 286.8--489.7 MHz band, as summarized in Table~\ref{tab:tab4}. This corresponds to a parasitic-capacitance correction of up to approximately 6 pF across different frequency bands. A comparison between the total effective inductance $L$ and the theoretical inductance of the secondary coil $L_{\mathrm{coil}}$ shows that the correction from the parasitic inductance is relatively small below 100 MHz but becomes dominant over $L_\mathrm{coil}$ at high-frequency bands. The parasitic inductance is about \(25~\mathrm{nH}\) in high frequency bands, thereby limiting the maximum resonant frequency to about 500 MHz with the present PCB design. The simplified LC model therefore reproduces the observed frequency tuning range and provides the order of magnitude of the parasitic capacitance and inductance. A full circuit model of how the parasitic capacitance and inductance depends on frequency and secondary coil geometry would be interesting but is beyond the scope of this work.

\begin{table}[h]
\caption{\label{tab:tab4}
A table summarizing the resonant-circuit parameters of the EOMs characterized in Fig.~4, including the frequency tuning range, effective crystal capacitance $\tilde{C}_{\mathrm{crystal}}$, effective inductance $L$, theoretical inductance of the secondary coil $L_{\mathrm{coil}}$, and the difference $\delta L = L-L_{\mathrm{coil}}$.
}
\begin{ruledtabular}
\begin{tabular}{lcccc}
Frequency (MHz) & $\tilde{C}_{\mathrm{crystal}}$ (pF) & $L$ (nH) & $L_{\mathrm{coil}}$ (nH) & $\delta L$ (nH) \\
$17.7\text{--}20.7$ & 9.3 & 2949 & 3080 & -131 \\
$50.0\text{--}67.4$   & 8.0  & 632  & 637 & -5  \\
$95.7\text{--}135.6$  & 6.3  & 193  & 168 & 25  \\
$161.0\text{--}233.5$ & 5.7  & 71   & 41 & 30\\
$247.3\text{--}407.5$ & 3.4  & 36   & 10 & 26\\
$286.8\text{--}489.7$ & 3.0  & 28   & 4 & 24\
\end{tabular}
\end{ruledtabular}

\end{table}

\section{Sideband generation at 461~MHz and RF-induced heating}
\label{app:461MHz}

\begin{figure}
    \centering
    \includegraphics[     width=\linewidth,     trim=0pt 10pt 0pt 20pt,     clip ]{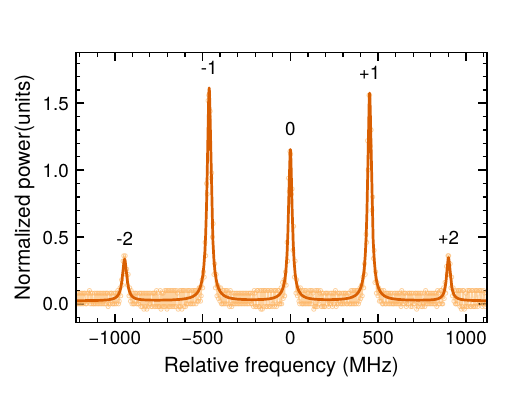}
    \caption{Optical sideband spectrum generated by the
    PCB-based resonant EOM driven at 461~MHz with an RF power
    of 35.6~dBm, measured using a scanning Fabry--P\'erot
    interferometer.
    The carrier ($0$), first-order sidebands ($\pm 1$),
    and second-order sidebands ($\pm 2$) are resolved at
    frequency offsets of $0$, $\pm 461$, and $\pm 922$~MHz,
    respectively.
    Open circles represent the measured transmission signal,
    and the solid curve shows the fit.
    The frequency axis is referenced to the carrier.}
    \label{fig:461MHz_sidebands}
\end{figure}

To demonstrate optical modulation at a frequency relevant to $^{39}$K laser cooling,
we measured the sideband spectrum with the EOM driven at 461~MHz using a 780-nm laser which is close to the potassium wavelength of 767 nm. This modulation frequency corresponds to the
ground-state hyperfine splitting of $^{39}$K.
Fig.~6 shows the FPI transmission spectrum obtained at an RF input power of 35.6~dBm
(approximately 3.63~W). The ratio of the first-order sideband to the carrier reaches about 1.4:1, which is substantially higher than the value of 0.8:1 typically required for laser cooling of $^{39}$K~\cite{Landini2012pra}.

At this RF input power, the resonant frequency shifts by approximately -5~MHz, which remains smaller than the resonance linewidth of the 461~MHz configuration.
The measured temperature dependence in Fig.~3(b) corresponds to a fractional temperature coefficient of approximately $-0.025\%/^{\circ}\mathrm{C}$. Assuming a comparable fractional temperature coefficient
for the 461~MHz configuration, the observed frequency shift implies an effective temperature rise of the resonant circuit of approximately $45\,^{\circ}\mathrm{C}$.
This estimate is based on an extrapolation of the ambient-temperature calibration rather than a direct measurement of the internal temperature, which is difficult as a thermal sensor would strongly disturb the high-frequency RF circuit. The inferred temperature rise is much larger than typical laboratory ambient-temperature fluctuations, supporting
RF-induced self-heating as the dominant origin of the observed high-power resonance shift.

Nevertheless, the separate 8-hour measurement at 228~MHz shown in Fig.~3(a) demonstrates that the sideband intensity remains stable after warm-up. This indicates that although RF heating can shift the resonance operating point, stable sideband generation can still be maintained once thermal equilibrium is established under the tested conditions.

\end{document}